\documentclass{article}
\usepackage{spconf,amsmath,graphicx}
\usepackage{epsfig}
\usepackage{amsfonts}
\usepackage{algorithm}
\usepackage{algorithmic}
\usepackage{multirow}
\usepackage{url}

\title{DiffSVC: A Diffusion Probabilistic Model for Singing Voice Conversion}
\name{Songxiang Liu$^{*1}$\thanks{$*$Equal contribution. Work done during internship in Tencent AI Lab}, Yuewen Cao$^{*1}$, Dan Su$^{2}$, Helen Meng$^{1}$}
\address{Author Affiliation(s)}

\address{
 $^1$Human-Computer Communications Laboratory,
 The Chinese University of Hong Kong\\
  $^2$Tencent AI Lab}
\begin{document}
%
\maketitle
\begin{abstract}
Singing voice conversion (SVC) is one promising technique which can enrich the way of human-computer interaction by endowing a computer the ability to produce high-fidelity and expressive singing voice.
In this paper, we propose DiffSVC, an SVC system based on denoising diffusion probabilistic model. DiffSVC uses phonetic posteriorgrams (PPGs) as content features. A denoising module is trained in DiffSVC, which takes destroyed mel spectrogram produced by the diffusion/forward process and its corresponding step information as input to predict the added Gaussian noise. We use PPGs, fundamental frequency features and loudness features as auxiliary input to assist the denoising process. Experiments show that DiffSVC can achieve superior conversion performance in terms of naturalness and voice similarity to current state-of-the-art SVC approaches.
\end{abstract}
\section{Introduction}
\label{sec1:intro}


Singing plays an important role in human daily life, including information transmission, emotional expression and entertainment.
The technology of singing voice conversion (SVC) aims at converting the voice of a singing signal to a voice of a target singer without changing the underlying content and melody. Endowing machine with the ability to produce high-fidelity and expressive singing voice provides new ways for human-computer interaction. SVC is among the possible ways to achieve this.

Most recent SVC systems train a content encoder to extract content features from a source singing signal and a conversion model to transform content features to either acoustic features or waveform. One class of SVC approaches jointly trains the content encoder and the conversion model as an auto-encoder model \cite{nachmani2019unsupervised,deng2020pitchnet}. Another class of SVC approaches separately trains the content encoder and the conversion model. These approaches train an automatic speech recognition (ASR) model as the content encoder. The ASR model can be an end-to-end model, as in \cite{polyak2020unsupervised,liu2020fastsvc} or a hybrid HMM-DNN model, as in \cite{li2020ppg}. The conversion model can be the generator in a generative adversarial network (GAN) \cite{polyak2020unsupervised,liu2020fastsvc}, which directly generates waveform from content features; or a regression model, which transforms content features to spectral features (e.g., mel spectrograms), and adopts an additionally trained neural vocoder to generate waveform.
In this paper, we focus on the latter class and devote to introducing the recently-emerged diffusion probabilistic modeling into the conversion model.

Diffusion probabilistic models, or diffusion models for brevity, are a class of promising generative models \cite{sohl2015deep}. It has been demonstrated that diffusion models are capable of achieving state-of-the-art performance on generative modeling for natural images \cite{NEURIPS2020_4c5bcfec, song2021denoising} and raw audio waveform \cite{kong2020diffwave,chen2020wavegrad}.
There are two processes in a diffusion model, i.e., the diffusion/forward process and the reverse process. The diffusion/forward process is a Markov chain with fixed parameters, which converts the complicated data into isotropic Gaussian distribution by adding Gaussian noise gradually. The reverse process is also a Markov chain, which restores the data structure from Gaussian noise in an iterative manner. One merit of diffusion models is that they can be efficiently trained by optimizing the evidence lower bound (ELBO), also known as variational lower bound (VLB), of the data likelihood. In \cite{NEURIPS2020_4c5bcfec}, one certain parameterization trick can convert the learning process of diffusion models into a regression problem. We follow this setting in this paper.

We propose DiffSVC, which is an SVC system based on diffusion model. We use an ASR acousic model to extract phonetic posteriorgrams (PPGs) from singing signals as the content features. A diffusion model is trained to recover mel spectrograms iteratively from Gaussian noise, conditioning on content, melody and loudness features. We show that DiffSVC can achieve superior conversion performance in terms of naturalness and voice similarity, compared with  current state-of-the-art SVC approaches. The contributions of our work include: (1) To the best of our knowledge, the proposed DiffSVC system is the first SVC system using the diffusion probabilistic model. We show that diffusion model can be effectively adopted for the SVC task; and (2) DiffSVC achieves better conversion performance in terms of naturalness and voice similarity than previous SVC systems.

The rest of this paper is organized as follows: Section~\ref{sec2} presents related work. Section~\ref{sec3} introduces the diffusion probabilistic model. Details of the proposed DiffSVC system are presented in Section~\ref{sec4}. Experimental results are shown in Section~\ref{sec5} and Section~\ref{sec6} concludes this paper.

 
\section{Related work}
\label{sec2}

\subsection{Singing Voice Conversion}

Current SVC approaches can be categorized into two classes, i.e., parallel SVC and non-parallel SVC, in terms of whether parallel training data is required. Most initial attempts for SVC are within the parallel SVC paradigm. These approaches model parallel training samples using statistical methods, such as Gaussian mixture model (GMM)-based many-to-many eigenvoice conversion \cite{toda2007one}, direct waveform modification based on spectrum difference \cite{kobayashi2014statistical,kobayashi2015statistical}. GAN-based parallel approach has also been proposed to improve conversion performance \cite{sisman2019singan}. Since parallel SVC approaches require parallel data, which is expensive to collect, for the training process, researchers have investigated many non-parallel SVC approaches. Auto-encoder based on the WaveNet \cite{oord2016wavenet} structure has been used for unsupervised SVC in \cite{nachmani2019unsupervised}, which can convert among singers appeared in the training set. This approach adopts an adversarial speaker classifier to disentangle singer information from the encoder output. To further improve this method, an additional pitch adversarial mechanism is added to remove pitch information from the encoder output in \cite{deng2020pitchnet}. Variational auto-encoder (VAE) \cite{luo2020singing}, variational autoencoding Wasserstein GAN (VAW-GAN) \cite{lu2020vaw}, and phonetic posteriorgram (PPG) models \cite{li2020ppg} are also investigated for non-parallel SVC. Very recently, the combination of a PPG model and a waveform generator achieves promising SVC performance \cite{polyak2020unsupervised,liu2020fastsvc}.

\subsection{Diffusion Probabilistic Models}

There are two streams of efforts, pushing the research on diffusion probabilistic models forward. One is research on the score matching model, which originates from \cite{JMLR:v6:hyvarinen05a}, where the problem of data density estimation is simplified into a score matching problem (the estimation of the gradient of the data distribution density). The other is research on denoising diffusion probabilistic model, which originates from \cite{sohl2015deep}, where a diffusion Markov chain is used to destroy data structure into Gaussian noise and another reverse process is used to generate data from Gaussian noise. Recently, great progress has been made in these two streams. In \cite{NEURIPS2020_4c5bcfec}, the diffusion model has been able to generate high-quality images. Slightly later, the diffusion model also makes it to generate high-quality audio samples \cite{kong2020diffwave,chen2020wavegrad}. On the other hand, score matching model has also been proved to be capable of generating high-resolution images using a neural network to estimate the log gradient of the target data probabilistic density function \cite{song2019_3001ef25}. Afterwards, a unified framework is proposed in \cite{song2021scorebased}, which generalizes and improves previous work on score matching model through the lens of stochastic differential equations (SDEs). The diffusion model in \cite{NEURIPS2020_4c5bcfec} and the score matching model in \cite{song2019_3001ef25} have been shown to be special cases under such framework.

\begin{figure}[t]
	\centering
	\includegraphics[width=0.5\textwidth]{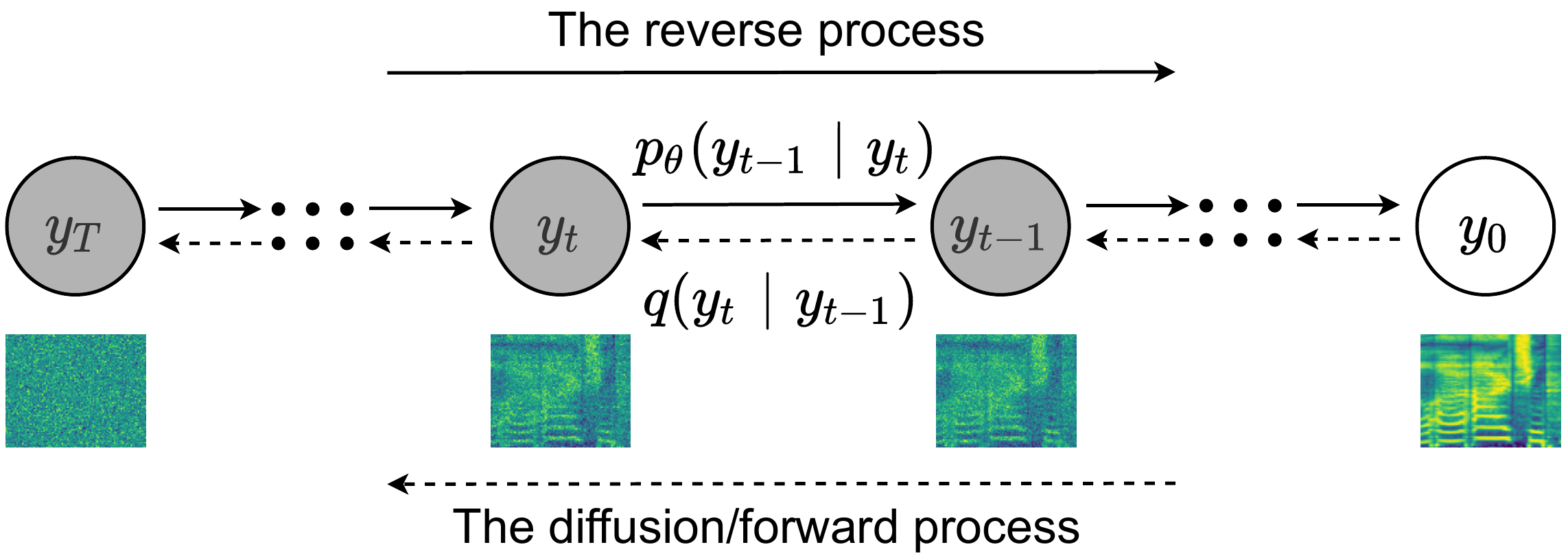}
	\caption{Graphical model for the reverse process and the diffusion/forward process in a diffusion probabilistic model. The diffusion/forward process (dotted arrow) gradually converts data into noise within finite steps. The reverse process (solid arrow) is a parametric procedure which attempts to restore data structure from noise.}
	\label{fig:diff_process}
\end{figure}

\begin{figure*}[t]
	\centering
	\includegraphics[width=16cm]{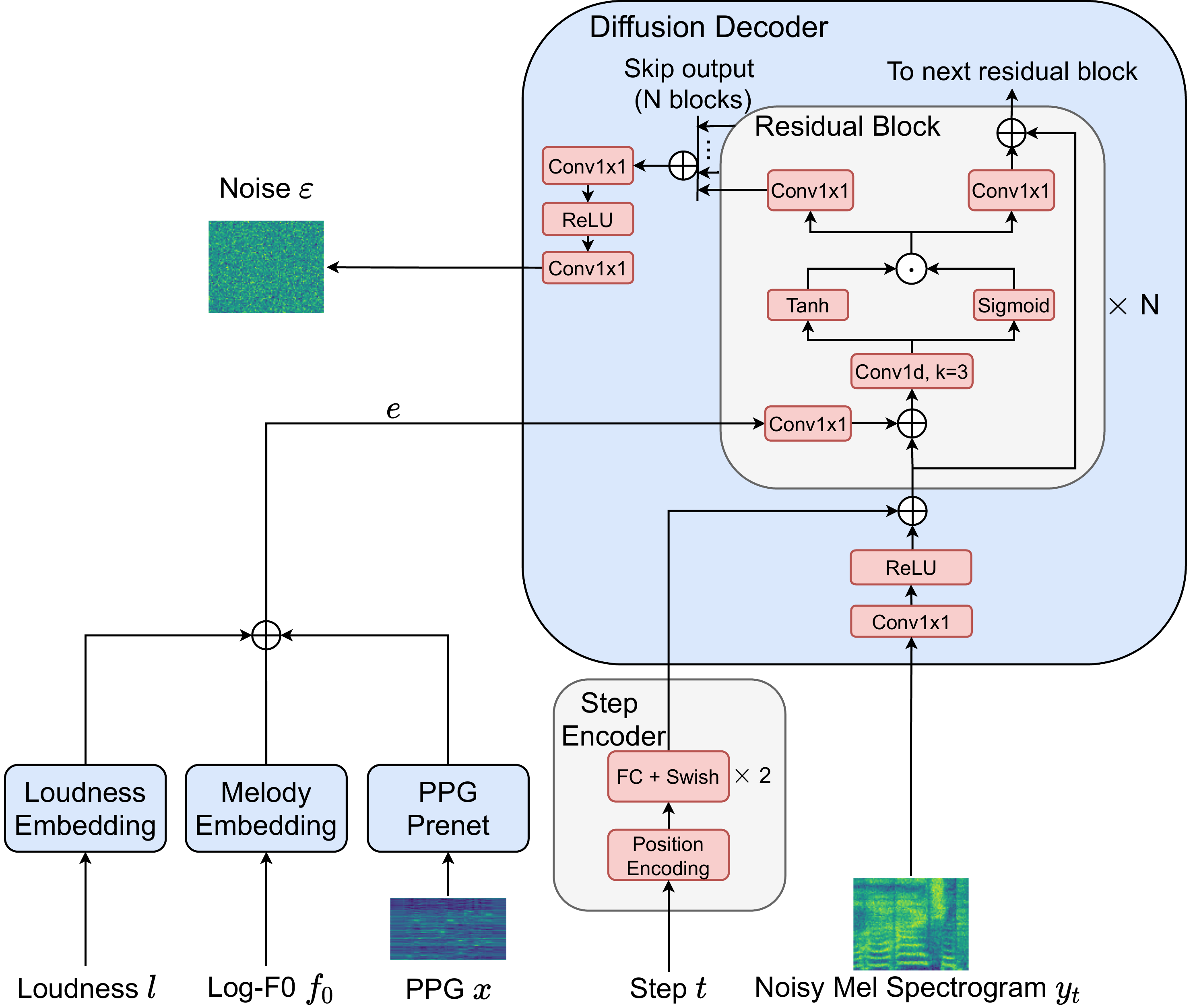}				
	\caption{Schematic diagram of the conversion model in the proposed DiffSVC system, based on the diffusion probabilistic model.}
	\label{fig:sys_overview}
\end{figure*}

\section{Diffusion Probabilistic Model}
\label{sec3}

In this section, we introduce the denoising diffusion probabilistic model (or diffusion model in short) \cite{sohl2015deep}. We denote the data as $y_0\sim q(y_0)$. A diffusion model is a latent variable model with the form $p_{\theta}(y_0) := \int p_\theta(y_{0:T})dy_{1:T}$, where $y_1, ..., y_T$ are latent variables with the same dimensionality as the data $y_0$ and $T$ is the total number of diffusion steps. As shown in Fig.~\ref{fig:diff_process}, a diffusion model contains two processes -- the diffusion/forward process and the reverse process. In the remaining part of this section, we first introduce the two processes in a diffusion model and then discuss the training and sampling algorithms.

\subsection{Diffusion/Forward Process}
The diffusion/forward process in a diffusion model is modeled as a Markov chain that gradually adds small noise to the data until the data structure is totally destroyed at step $T$. That is, the diffusion process gradually whitens the data into Gaussian noise in $T$ diffusion steps. The chain transitions are often modeled as conditional Gaussian transitions according to a deterministic noise schedule $\beta_1, ..., \beta_T$, which are set to satisfy $\beta_1 < \beta_2 < ... < \beta_T$. The joint probability of the latent variables is:
\begin{equation}
    q(y_{1:T}|y_0) := \prod_{t=1}^{T} q(y_t|y_{t-1}),
\end{equation}
where $q(y_t|y_{t-1}) := \mathcal{N}(y_t;\sqrt{1-\beta_t}y_{t-1}, \beta_t I)$. The mean term $\sqrt{1-\beta_t}$ gradually decays the destroyed data towards origin, and the variance term $\beta_t$ incorporates stochasticity into the process by adding small noises, for $t\in\{1, ..., T\}$.

\subsection{Reverse Process}
The reverse process in a diffusion model is a parameterised process which gradually learns to restore data structure from white noise along another Markov chain. When $\beta$'s of the diffusion/forward process are small, both the diffusion and reverse process have the same functional form \cite{sohl2015deep}, i.e., Gaussian transitions. Starting at isotropic Gaussian $p(y_T)=\mathcal{N}({y_T; 0, I}$), the joint distribution $p_\theta(y_{0:T})$ along the reverse process satisfies:
\begin{equation}
    p_\theta(y_{0:T}) := p(y_T)\prod_{t=1}^T p_\theta(y_{t-1}|y_t),
\end{equation}
where the transition probability $p_\theta(y_{t-1}|y_t)$ is parameterized as $\mathcal{N}(y_{t-1};\mu_\theta(y_t, t), \sigma_\theta(y_t, t)^2 I)$ with shared parameter $\theta$ among all the $T$ transitions. The training process of a diffusion model is to learn the parameter $\theta$, such that $p_\theta(y_{t-1}|y_t)$ can eliminate the Gaussian noise (i.e., denoise) added during the diffusion/forward process.

\subsection{Training Diffusion Model}
The goal of training a diffusion model is to maximize the model likelihood $p_{\theta}(y_0) := \int p_\theta(y_{0:T})dy_{1:T}$,  given a training set $\{y_0^m\}_{m=1}^M$ with $M$ samples. However, the likelihood is intractable to compute in general because of the integration operation in a very high-dimensional latent space. Thus, its evidence lower bound (ELBO), which is also known as variational lower bound (VLB), is used to train the model:
\begin{equation} \label{eq:elbo}
\mathbb{E}_{q(y_0)}[\log p_\theta(y_0)]\geq \underbrace{\mathbb{E}_{q(y_{0:T})}[\log\frac{p_\theta(y_{0:T-1}|y_T)p(y_T)}{q(y_{1:T}|y_0)}]}_{:=ELBO}
\end{equation}
We refer readers to \cite{sohl2015deep} for a complete proof of Eq.~(\ref{eq:elbo}).

Thanks to the property of Gaussian transitions in the diffusion process, the noisy data $y_t$ (i.e., destroyed data) at any step $t\in\{1, ..., T\}$ can be readily sampled given a data sample $y_0$:
\begin{equation}
    q(y_t|y_0) = \mathcal{N}(y_t;\sqrt{\bar\alpha_t}y_0,\sqrt{1-\bar\alpha_t}I),
\end{equation}
where $\alpha_t:=1-\beta_t$ and $\bar\alpha_t=\prod_{s=1}^t\alpha_s$.
Following the reparameterization trick presented in \cite{NEURIPS2020_4c5bcfec}, the learning process becomes a regression problem, where a neural network (parameterized with $\theta$) is trained, which takes the noisy data $y_t$ and the step variable $t$ as input to predict the  corresponding noise $\epsilon$. Using a simplified variant of the ELBO in Eq.~(\ref{eq:elbo}) as in \cite{NEURIPS2020_4c5bcfec}, the negative ELBO becomes:
\begin{equation} \label{eq:simplified-elbo}
    -ELBO=\mathbb{E}_{y_0, t}[||\epsilon - \epsilon_\theta(\sqrt{\bar\alpha_t}y_0+\sqrt{1-\bar\alpha_t}\epsilon, t)||_2^2],
\end{equation}
where $\epsilon\sim\mathcal{N}(0,I)$, $y_0\sim q(y_0)$, and $t$ is uniformly sampled from $1,...,T$. In this paper, we use Eq.~(\ref{eq:simplified-elbo}) to train the diffusion model.

\subsection{Sampling with Langevin Dynamics}

Given the parameterized reverse process with the well-learned parameter $\theta$, the sampling process (i.e., the generative process) is to first sample a Gaussian noise $y_T \sim \mathcal{N}(0, I)$, and then iteratively sample $y_{t-1}\sim p_\theta(y_{t-1}|y_t)$ for $t=T, T-1, ..., 1$ along the reverse process, according to Langevin dynamics:
\begin{equation} \label{eq:sample}
    y_{t-1} = \frac{1}{\sqrt{\alpha_t}}(y_t - \frac{1-\alpha_t}{\sqrt{1-\bar{\alpha}_t}}\epsilon_\theta(y_t, t))+\sigma_t z,
\end{equation}
where $\sigma_t^2=\frac{1-\bar{\alpha}_{t-1}}{1-\bar{\alpha}_t}\beta_t$, $z\sim\mathcal{N}(0,I)$ for $t>1$ and $z=0$ for $t=1$. The final $y_0$ is the generated data.

\section{DiffSVC}
\label{sec4}

In this section, we describe how we introduce the diffusion model to SVC and present details of the proposed DiffSVC system.
We regard the content, melody and loudness information as the most important components in a singing signal for the task of singing voice conversion. In DiffSVC, we train a Deep-FSMN (DFSMN)-based ASR acoustic model \cite{zhang2018deep} to extract PPGs as the content features with forced-aligned audio-text speech data. We train the ASR model with frame-wise cross-entropy loss, where the ground-truth labels are phonemes (i.e., initials and finals with tone) for Mandarin Chinese SVC.

We introduce the concept of denoising diffusion modeling into the conversion model in DiffSVC. The overall architecture is depicted in Fig.~\ref{fig:sys_overview}. During the training process, DiffSVC aims at predicting the isotropic Gaussian noise $\epsilon$ from the noisy mel spectrogram $y_t$ and the step variable $t$, leveraging additional information from PPG $x$, logarithmic fundamental frequency feature (Log-F0) $f_0$ and loudness feature $l$. We introduce the details in the following subsections.

\subsection{PPG, Log-F0 and Loudness}
The PPG prenet adopts a simple fully-connected (FC) layer on the PPG input. DiffSVC represents the melody of a singing signal using its fundamental frequency contour in logarithmic scale. A-weighting mechanism of a singing signal's power spectrum is adopted to compute loudness features, following \cite{liu2020fastsvc}. We process the Log-F0 features and loudness features in the same way. The Log-F0 features and loudness features are first quantized into 256 bins, and then go through a melody embedding lookup table and a loudness embedding lookup table respectively. 
The processed PPG features, Log-F0 features and loudness features are added elementwisely to obtain the conditioner $e$, which is token as additional input to the diffusion decoder.

\subsection{Diffusion Modeling}

\textbf{Step Encoding}. The diffusion step variable $t$ is an important input to a diffusion model, because it indicates the amount of noise added to the noisy mel spectrogram $y_t$. Following the setting in \cite{kong2020diffwave}, we first convert integral-valued $t$ into an 128-dimensional vector $t_{emb}$ with a sinusoidal position encoding as:
\begin{equation}
    [\sin(10^{\frac{0\times4}{63}}t), ..., \sin(10^{\frac{63\times4}{63}}t), \cos(10^{\frac{0\times4}{63}}t), ..., \cos(10^{\frac{63\times4}{63}}t)]
\end{equation}
Then we further process $t_{emb}$ with two FC layers and Swish activation function \cite{ramachandran2017searching}. \\
\textbf{Diffusion Decoder}. The diffusion decoder is trained to predict the noise $\epsilon$ added by the diffusion process from the noisy mel spectrogram $y_t$ and the step variable $t$, taking auxiliary conditioners $e$ obtained from PPG $x$, Log-F0 $f_0$ and loudness $l$.
Following \cite{kong2020diffwave}, the diffusion decoder in DiffSVC adopts a bidirectional residual convolutional architecture proposed in \cite{oord2016wavenet} with a slight difference. We use dialtion rate of 1 since we work on mel spectrograms instead of raw waveform. The receptive field of the diffusion decoder is sufficiently large with a dilation rate of 1. The diffusion decoder first conducts an one-dimensional convolutional operation with kernel-size 1 (Conv1x1) on the noisy mel spectrogram and then adopts the ReLU activation on the output. The output of the step encoder is added to every time step of the mel spectrogram feature maps, and then fed into $N$ residual blocks. The conditioner $e$ is imported into every residual block with separate Conv1x1 layers, whose output is elementwisely added with the mel spectrogram and step features. Then the gated mechanism introduced in \cite{oord2016wavenet} are used to further process the feature maps. We sum the skip connections from all the $N$ residual layers and further process with two Conv1x1 layers interleaved with a ReLU activation to get the diffusion decoder output. \\
\textbf{Training and conversion} We summarize the training procedure of the conversion model in DiffSVC in Algorithm~\ref{alg:training} and the conversion procedure in Algorithm~\ref{alg:conversion}, respectively.

\begin{algorithm}[h]
\caption{Training procedure of the conversion model in DiffSVC.}
\begin{algorithmic}[1]
\label{alg:training}
\REQUIRE The conversion model $\epsilon_\theta(\cdot)$; the training set $\mathcal{D}_{train}=\{(x, f0, l, y_0)\}_{m=1}^M$; $N_{iter}$ training iterations.
\FOR{$i=1,2,...,N_{iter}$}
\STATE Sample $(x, f_0, l, y_0)$ from $\mathcal{D}_{train}$;
\STATE $\epsilon\sim\mathcal{N}(0,I)$;
\STATE Sample $t\sim$ Uniform$(\{1, \cdots, T\})$;
\STATE Take gradient descent step on \\ \quad$\nabla_\theta||\epsilon-\epsilon_\theta(\sqrt{\bar\alpha_t}y_0+\sqrt{1-\bar\alpha_t}\epsilon, t, x, f_0, l))||_2^2$;
\ENDFOR
\RETURN $\epsilon_\theta(\cdot)$;
\end{algorithmic}
\end{algorithm}

\begin{algorithm}[h]
\caption{Conversion procedure of DiffSVC.}
\begin{algorithmic}[1]
\label{alg:conversion}
\REQUIRE The trained conversion model $\epsilon_\theta(\cdot)$; one testing sample $(x, f0, l)$.
\STATE Sample $y_T\sim\mathcal{N}(0,I)$;
\FOR{$t=T,T-1,...,1$}
\STATE Sample $(x, f_0, l, y_0)$ from $\mathcal{D}$;
\STATE $\epsilon\sim\mathcal{N}(0,I)$;
\IF{$t>1$}
\STATE Sample $z\sim\mathcal{N}(0, I)$;
\ELSE
\STATE $z=0$;
\ENDIF
\STATE $y_{t-1} = \frac{1}{\sqrt{\alpha_t}}(y_t - \frac{1-\alpha_t}{\sqrt{1-\bar{\alpha}_t}}\epsilon_\theta(y_t, t, x, f_0, l))+\sigma_t z$, as in Eq.~\ref{eq:sample};
\ENDFOR
\RETURN $y_0$;
\end{algorithmic}
\end{algorithm}

\section{Experiments}
\label{sec5}

We evaluate the proposed DiffSVC by conducting any-to-one SVC tasks, where  the singing voice of an arbitrary source singer is converted to the target singer's. In this section, we first introduce the corpus for the experiments and the data pre-processing methods. Then the implementation details of both the baseline approaches and proposed approach are presented. Finally, we present the evaluation results.

\subsection{Dataset and preprocessing}
\label{data_prep}
The PPG extractor is trained using an internal Mandarin Chinese ASR corpus, which contains around 20k hours speech data recorded by thousands of speakers. The PPG features have dimensionality of 218. The conversion model in DiffSVC is trained with an internal dataset, which contains 14 hours singing data recorded by a female voice talent. The audio format is 16-bit PCM with a sample rate of 24 kHz. Mel spectrograms have 80 frequency bins, and are computed with 1024-point fast Fourier transform (FFT), 1024-point window size and 240-point hop size (i.e., 10 ms). Mel spectrograms are min-max normalized to the range [-1, 1]. Fundamental frequency (F0) values are computed with hop size of 240. To make the F0 computation robust, we use three F0 estimators, i.e., DIO \cite{morise2009fast}, REAPER \footnote{[Online] \url{https://github.com/google/REAPER}} and SPTK \cite{sptk}. The F0 values take the median results by this three F0 estimators. The FFT size, window size and hop size for computing loudness features are 2048, 2048 and 240, respectively.

\subsection{Implementation details}
\label{implementation}
The output size of the PPG prenet, the melody embedding size and loudness embedding size are all 256. The number of bins for quantizing Log-F0 and loudness is 256. The step encoder has the same hyper-parameter setting as in \cite{kong2020diffwave}. We use 20 residual layers in a diffusion decoder, where the number of convolution channels are 256. The total number of diffusion steps is 100 (i.e., T = 100). The noise schedule $\beta$'s are set to linearly spaced from $1\times 10^{-4}$ to 0.06. We train models with the ADAM optimizer \cite{kingma2014adam} with a constant learning rate of $0.0002$.

We train a Hifi-GAN model \cite{hifigan} as vocoder to convert mel spectrograms to the raw waveform. The official implementation\footnote{[Online] \url{https://github.com/jik876/hifi-gan}} (the ``config\_v1.json'' configuration) is used with a slight change. Since we use 24 kHz audio samples and the hop-size for computing mel spectrogram is 240, we factorize the upsample rate as $240=8\times5\times3\times2$. To avoid possible checkerboard artifacts caused by the ``ConvTranspose1d" upsampling layer, we use temporal nearest interpolation layer followed by a one-dimensional convolutional layer as the upsampling operation, as suggested in \cite{odena2016deconvolution}. The kernel sizes we use are [15,15,5,5].

\subsection{Comparisons}

We compare the proposed DiffSVC system with three state-of-the-art SVC approaches. To make the comparison fair, all models use the same DFSMN-based PPG extractor (as introduced in Section~\ref{sec4}) to compute PPG features as the content representations. Fundamental frequency ($f_0$) values are computed using the robust ensemble method introduced in Section~\ref{data_prep}.
The details of their implementations are presented as follows.  \\
\textbf{BLSTM-SVC} This model is proposed by \cite{chen2019singing}. We make slight modifications to make the comparison fair: (1) We use mel spectrogram and Log-F0 as the acoustic features instead of the conventional vocoder parameters such as mel cepstral (MCEP) features and aperiodicity (AP) features; and (2) we use the Hifi-GAN vocoder introduced in Section~\ref{implementation} for high-fidelity waveform generation. We reimplemented the 512N model in \cite{chen2019singing} since it shows better conversion performance according to the experimental results in the orginal paper.  \\
\textbf{Seq2seq-SVC} This model is proposed by \cite{li2020ppg}, where a sequence-to-sequence (seq2seq) based model using GMM-based attention mechanism is trained to predict mel spectrograms from PPGs. Slight modifications are also made to make the comparison fair: (1) we do not use the Mel encoder and the singer confusion module because we focus on single-speaker conversion model; (2) we use the Hifi-GAN vocoder introduced in Section~\ref{implementation} for converting mel spectrograms into waveform, instead of using a WaveRNN vocoder\cite{kalchbrenner2018efficient}. \\
\textbf{FastSVC} This model is presented in \cite{liu2020fastsvc}, which uses a generator in a GAN model to convert content features directly into waveform, and is trained by combining a multi-scale spectral loss and an adversarial loss. In the original paper, the content features used by the authors have a hop-size of 320, while the PPGs used in this paper have a hop-size of 240. Therefore, we change the upsampling rates of the generator to [5,4,4,3] from the original [4,4,4,5]. To further adapt FastSVC for 24 kHz audio synthesizing (originally, FastSVC works on 16 kHz audio), we add the multi-period discriminator (MPD) \cite{hifigan} as additional discriminator. These changes make FastSVC achieving better conversion performance on 24 kHz audio samples.

\subsection{Evaluations}
Subjective evaluations in terms of both the naturalness and voice similarity of converted speech are conducted. We use the 5 point Likert scale for testing with mean opinion score (MOS) (1-bad, 2-poor, 3-fair, 4-good, 5-excellent). In the MOS tests for evaluating naturalness, each group of stimuli contains recording samples from the target singer, which are randomly shuffled with the samples generated by the four comparative approaches, before they are presented to raters. In the MOS similarity tests, converted speech samples are directly compared with the recording samples of the target singer. 
10 utterances from an evaluation set are presented for each system. 
We invited raters to participate in the evaluations in a quiet room and they were asked to use headphones during the tests. The raters were allowed to replay each sample as many times as necessary and change their ratings of any sample before submitting their results.
The MOS results are demonstrated in Table~\ref{tab:1}. We can see that DiffSVC achieves significantly better performance than the other three compared systems (i.e., BLSTM-SVC, Seq2seq-SVC and FastSVC), in terms of  both naturalness and voice similarity. Audio samples can be found online\footnote{[Online] \url{https://liusongxiang.github.io/diffsvc/}}.

\begin{table}[]
\caption{Subjective and objective evaluation results. ``Nat." is for ``Naturalness" and ``Sim." is for ``Voice Similarity", repectively. Up arrows mean ``higher the better". Down arrows mean ``lower the better".}
\vspace{0.3cm}
\resizebox{0.48\textwidth}{!}{
\begin{tabular}{|c|c|c|c|c|}
\hline
\multirow{2}{*}{System}&\multicolumn{2}{c|}{Subjective}&\multicolumn{2}{c|}{Objective} \\ \cline{2-5} 
&\multicolumn{1}{c|}{Nat. ($\uparrow$)}&\multicolumn{1}{c|}{Sim.($\uparrow$)}&MCD ($\downarrow$)&FPC($\uparrow$)\\ \hline
Recordings  & 4.60$\pm$0.09 & \multicolumn{1}{c|}{-}                   & \multicolumn{1}{c|}{-} & \multicolumn{1}{c|}{-} \\ \hline
BLSTM-SVC  & 2.37$\pm$0.15 & 3.67$\pm$0.15   & 6.424   & 0.781      \\ \hline
Seq2seq-SVC & 2.77$\pm$0.17 & 3.57$\pm$0.15   & 7.175   & 0.885      \\ \hline
FastSVC     & 3.83$\pm$0.13 & 4.17$\pm$0.13   & 6.422   & \textbf{0.904} \\ \hline
DiffSVC     & \textbf{3.97$\pm$0.14}  & \textbf{4.67$\pm$0.10}  & \textbf{6.307} & 0.902                  \\ \hline
\end{tabular}}
\label{tab:1}
\end{table}

We use mel-cepstrum distortion (MCD) and F0 Pearson correlation (FPC) as metrics for objective evaluations. We extract the mel-cepstrum of the generated audio samples and the ground-truth audio samples from the target singer for MCD evaluation. For the FPC evaluation, we extract the F0 from  the generated audio samples and the recordings from the source singer, since our goal is to convert the voice of a singing signal to the target singer's voice without changing the underlying content and melody. The DTW algorithm is adopted to align the predicted acoustic features towards the natural ones.  
The objective results are presented in Table~\ref{tab:1}. We can see that, among the four compared systems, DiffSVC achieves the best MCD result (6.307). In terms of FPC, the FastSVC system achieves the best result (0.904). DiffSVC achieves slightly lower FPC result than FastSVC, with an FPC value of 0.902 and an 0.2\% relative drop.

Combining the experimental results, we can safely say that DiffSVC achieves superior singing voice conversion performance than previous SVC approaches in terms of naturalness and voice similarity of the converted signal.

\section{Conclusion}
\label{sec6}

In this paper, we have presented DiffSVC. To the best of our knowledge, DiffSVC is the first SVC approach using a diffusion probabilistic model to generate acoustic features (e.g., mel spectrogram) from content features (e.g., PPG). 
Experiments demonstrate that the diffusion model can be effectively adopted for the SVC task. Moreover, DiffSVC can achieve superior conversion performance in terms of naturalness and voice similarity than current state-of-the-art SVC systems, according to the subjective and objective evaluations. 
In the work, we focus on any-to-one SVC task. Future work includes further extending DiffSVC for any-to-many/any SVC. Investigation on training a DiffSVC model under low-data-resource scenario is also deserved research effort.

\bibliographystyle{IEEEbib}
\bibliography{strings,refs}

\end{document}